\documentclass[iop]{emulateapj}
\usepackage{subfigure}
\usepackage{wrapfig}

\def\lsim{\mathrel{\rlap{\lower4pt\hbox{\hskip1pt$\sim$}}
    \raise1pt\hbox{$<$}}}                
\def\gsim{\mathrel{\rlap{\lower4pt\hbox{\hskip1pt$\sim$}}
    \raise1pt\hbox{$>$}}}                

\shorttitle{Imaging and wavefront sensing with IRAPDs}
\shortauthors{Baranec et al.}
\begin{document}
\title{High-speed imaging and wavefront sensing with an infrared avalanche photodiode array}

\author{Christoph Baranec\altaffilmark{1}, Dani Atkinson\altaffilmark{1}, Reed Riddle\altaffilmark{2}, Donald Hall\altaffilmark{1}, Shane Jacobson\altaffilmark{1}, Nicholas M. Law\altaffilmark{3}, and Mark Chun\altaffilmark{1}}

\altaffiltext{1}{Institute for Astronomy, University of Hawai`i at M\={a}noa, Hilo, HI 96720-2700, USA; baranec@hawaii.edu}
\altaffiltext{2}{Division of Physics, Mathematics, and Astronomy, California Institute of Technology, Pasadena, CA, 91125, USA}
\altaffiltext{3}{Department of Physics and Astronomy, University of North Carolina at Chapel Hill, Chapel Hill, NC 27599-3255, USA}

\begin{abstract}
Infrared avalanche photodiode arrays represent a panacea for many branches of astronomy by enabling extremely low-noise, high-speed and even photon-counting measurements at near-infrared wavelengths. We recently demonstrated the use of an early engineering-grade infrared avalanche photodiode array that achieves a correlated double sampling read noise of  0.73 e$^{-}$ in the lab, and a total noise of 2.52 e$^{-}$ on sky, and supports simultaneous high-speed imaging and tip-tilt wavefront sensing with the Robo-AO visible-light laser adaptive optics system at the Palomar Observatory 1.5-m telescope. We report here on the improved image quality achieved simultaneously at visible and infrared wavelengths by using the array as part of an image stabilization control-loop with adaptive-optics sharpened guide stars. We also discuss a newly enabled survey of nearby late M-dwarf multiplicity as well as future uses of this technology in other adaptive optics and high-contrast imaging applications. 
\end{abstract}


\maketitle

\def \Kepler {\textit{Kepler}}
\section{Introduction}

Solid-state infrared detectors have made major contributions to our understanding of the universe over the past several decades \citep{Low2007}. Recent innovations in infrared avalanche photodiode (APD) detectors, wherein the avalanche gain of photo-generated electrons occurs within the HgCdTe substrate, have reduced the effective read noise of sizable pixel arrays to below the critical 1 e$^{-}$ threshold \citep{Feautrier2014, Finger2014}. When paired with correspondingly-low dark currents, there is potential for drastically improving the many current and future applications of infrared arrays in astronomy: e.g., infrared photon counting \citep{Beletic2013, Rauscher2015}, improving sky coverage of laser guide star adaptive optics (AO) systems using sharpened infrared tip-tilt stars \citep{Dekany2008, Wizinowich2014, McCarthy1998, Wang2008}, increasing sensitivity of pyramid wavefront sensors \citep{Peter2010} and interferometers, e.g., S. Guieu et al. (2015, in preparation), decreasing noise in post-coronagraphic and speckle nulling wavefront sensors in high-contrast systems \citep{Cady2013, Martinache2012} and improving temporal bandwidth and sensitivity for IR photometric observations \citep{Mereghetti2008, Rafelski2006}. To prove this maturing technology in a challenging observing environment, we demonstrated the use of a Selex ES Advanced Photodiode for High-speed Infrared Array (SAPHIRA) with the Robo-AO visible-light laser adaptive optics system \citep{Baranec2014} mounted to the robotic Palomar Observatory 1.5-m telescope \citep{Cenko2006}. During observations, the full 320 $\times$ 256 pixel SAPHIRA array was operated in 32-ouput mode at a 265 kHz pixel rate. This allowed the array to be read out (and recorded) at 100 frames per second. The position of a star in the infrared field was calculated and used to stabilize image displacement with a beam steering mirror in the Robo-AO system. In this paper, we describe the experimental setup that integrated a camera with a SAPHIRA detector with the Robo-AO system (Section~\ref{sec:exp}), describe initial results and delivered image quality including a pilot survey of very red nearby stars for multiplicity (Section~\ref{sec:results}), and detail future work and plans for the SAPHIRA technology (Section~\ref{sec:plans}). 

\section{Experimental Setup}
\label{sec:exp}

The Robo-AO system offers a flexible testing environment for new cameras and instruments requiring visible and infrared diffraction-limited capability. The system operates under Linux with the control software source code written in the C++ language \citep{Riddle2012}. The Robo-AO Cassegrain instrument package has two ports for external instruments: one visible port fed by a manually installed visible beamsplitter in front of the main EMCCD science camera (previously used with an eyepiece; \citealt{Baranec2012}), and another infrared port fed by transmission through a dichroic mirror passing $\lambda > 950$ nm and subsequent reflection off of a gold mirror \citep{Baranec2013}. Both external ports have an unvignetted field-of-view of 2\arcmin ~with a telecentric F/41 output.  A fast beam steering mirror is incorporated into the adaptive optics relay ahead of the visible-infrared dichroic mirror and is used for global tip-tilt correction of the science field.

\subsection{Infrared avalanche photodiode array camera}

In this experiment, we used an engineering-grade non-anti-reflection-coated Mark 3 Selex ES Infrared SAPHIRA detector \citep{Finger2014, Atkinson2014}. The detector was a metal organic phase epitaxy (MOVPE) HgCdTe avalanche photodiode array comprising 24 $\mu$m square pixels in a 320 $\times$ 256 format. The detector was located at the Robo-AO F/41 focus with a plate scale of 0\farcs079 and installed into a GL Scientific Stirling Cooler Cryostat which maintained an equilibrium temperature of 85 K. The cryostat was attached to the Robo-AO infrared port with a sliding interface plate to position the detector on the optical axis and shimmed to achieve optimal focus. A Mauna Kea Observatories H-band filter ($\overline{\lambda}$= 1.635 µm; \citealt{Tokunaga2002}) that also blocks longer wavelength radiation beyond the sensitivity of the detector was mounted inside the cryostat, in front of and in thermal contact with the detector array.

The SAPHIRA detector was controlled by a commercially available third-generation controller system produced by Astronomical Research Cameras (ARC), mounted adjacent to the cryostat. Clock voltages and detector readouts were provided by an ARC-32 Clock Driver Board and four eight-channel ARC-46 IR Video Boards, respectively. An ARC-22 Timing Board provided fiber optic communication with a PC and regulated timing within the controller. Alongside the standard ARC boards was a custom analog board that provided stable, low-noise supply and bias voltages, designed and developed at Australian National University and reproduced (and modified) for use with the SAPHIRA detector. The controller behavior was dictated by code loaded from the connected PC, and was written and compiled in Motorola DSP56000 assembly.

The SAPHIRA detector was operated with a detector bias voltage of 11.5V, corresponding to an avalanche gain of 22. Prior laboratory testing of the detector in a dark cryostat at this bias voltage \citep{Atkinson2014} showed an effective root-mean-square (RMS) read noise of 0.73 e$^-$ after avalanche gain for a single correlated double sampling (CDS) read and 72 e$^-$/s of dark current. When deployed on sky, we measured a total amount of noise of 2.52$\pm$0.18 e$^-$ per CDS read. Assuming the same read and dark noise as found in the lab, and a sky brightness of 13.7 mag/sq. arc sec (0.3 e$^-$ RMS), there is a remaining 2.2 e$^-$ RMS of noise that is not accounted for. While much of this can be attributed to instrument emissivity and the lack of baffling in front of the detector, a higher on-sky read or dark noise cannot be precluded.   

We observed an $m_H$ = 10.85 star, with an expected 212,000 photons/s expected at the telescope aperture \citep{Cohen1992}, and measured 25,000$\pm$2,400 photo-e$^-$/s leading to a total system throughput of 11.8$\pm$1.1$\%$. This is consistent with a throughput estimate of 10.1-11.8$\%$ based on the estimated throughput of the telescope of ~72$\%$ (two bare aluminum reflections), adaptive optics system in H band of 55$\%$ (from measured reflection and transmission data from all optical components), average in-band transmission of the H-band filter of 85$\%$, and the quantum efficiency of the SAPHIRA of 30-35$\%$ (limited by Fresnel reflection of uncoated HgCdTe).

\subsection{Integration with the Robo-AO adaptive optics system}

The device driver for the ARC PC interface card was incompatible with the Robo-AO Fedora 13 operating system necessitating a separate computer to host the SAPHIRA camera with inter-computer communications running over gigabit Ethernet, taking advantage of the multiple computer communication routines previously developed. The Robo-AO/SAPHIRA control software was adapted from the control software for the Robo-AO EMCCD camera. The lower level software integrated the SAPHIRA driver into the Robo-AO architecture to control basic functions (e.g., opening the camera connection, setting parameters, taking an image).  A second layer to the software created a generalized control system for all cameras between the hardware interface and the daemon control system common to all Robo-AO subsystems; this layer was modified to control the new functionality of the SAPHIRA detector.

The Robo-AO/SAPHIRA control software provided full control over the SAPHIRA detector for operations as a science detector and a tip-tilt sensor.  In both cases, the full array was read out at 100 frames per second, limited by the 265 kHz pixel rate of the ARC electronics. Reading of the array is a non-destructive process. To avoid possible non-linear response or the saturation of pixels the array needs to be reset, taking the same time as a read, well before the full-well depth of any pixel is exceeded. The array reset rate, measured as the number of frames to read prior to a reset, is uploaded to the ARC electronics as part of the camera configuration process. For this experiment we selected a fixed reset rate of 32 frames; this proved to be too long for the bright star in the experiment in Sections \ref{threeone}-2. In practice the reset rate should be tailored to accommodate the brightest source expected to be observed, with additional overhead for uncertainties. During an image acquisition sequence, both raw pixel reads and calibrated difference frames were recorded. The calibration of individual frames comprised subtracting a sky background, normalizing by a flat-field and applying a static hot-pixel mask (totaling approximately 3$\%$ of pixels) wherein hot-pixel values were replaced by a median of the surrounding eight pixel values. In the calibrated data, frames recorded immediately after an array reset include very negative values and were ignored by the tip-tilt system.

Observations with the SAPHIRA camera system required manually starting each sub-system of the observation sequence as opposed to being fully integrated into the Robo-AO robotic operations and queue system. When the SAPHIRA camera was used as a tip-tilt sensor, an initial image was taken with the camera with the high-order adaptive optics control loop operating which sharpened the instantaneous stellar point-spread-function (PSF). Once a tip-tilt reference star was identified in the field, an 8 $\times$ 8 pixel, 0\farcs63 $\times$ 0\farcs63 window ($\sim3\lambda$/D) centered on the star was defined in the tip-tilt configuration file. The tip-tilt compensation sub-system was started independent of other processes; as each calibrated output frame was recorded, the position of the star was calculated using a center-of-mass algorithm on the windowed pixels. The displacement of the star from the center of the pixel window was transmitted to the Robo-AO control computer. New fast steering mirror actuator position commands were calculated to re-center the star with a loop gain of 0.5 and were applied during the next cycle of the asynchronous 1.2 kHz high-order adaptive optics control loop. The latency of the tip-tilt compensation was dominated by the 10 ms read-time of the infrared array, followed by $\sim$1ms for inter-computer communication and $< 4 \mu$s for the frame calibration and center-of-mass calculation.

\section{Results}
\label{sec:results}

\subsection{Technical observations}
\label{threeone}

The SAPHIRA camera was paired and tested with the Robo-AO system on 2014 September 3, 08:13-09:12 UT. Through-telescope seeing was measured to be $\sim$1\farcs0 in a long-exposure open-loop image in the Sloan i'-band ($\overline{\lambda}$=765 nm) at the beginning of the testing period. To confirm the stability of the seeing measurements, we monitored the image width on the nearby Palomar 48-inch telescope; the seeing remained very steady, with an RMS of 0\farcs10 over the entire night. All observations reported here were 2 minutes in duration. Observations with the visible-light EMCCD camera were taken as a series of full-frame reads at the maximum rate of 8.6 frames per second in i'-band. The full width at half maximum (FWHM) were determined from a calculated best-fit 2D Gaussian, with random errors typically on the order of 0\farcs04. Strehl ratios were calculated by first simulating a perfect PSF and normalizing the peak intensity by the flux within a 3\farcs0 square aperture, accounting for 98$\%$ and 96$\%$ of the total energy in i'- and H-bands respectively. The peak of the stellar PSF was normalized to the flux within the same sized aperture and the ratio of the flux normalized peaks in both PSFs produced the Strehl ratio. Systematic errors on the Strehl ratio are due to pixel-grid alignment errors and not accounting for 100$\%$ of the scattered light in the stellar halo and were typically on the order of 10$\%$ of the calculated value.

\begin{figure*}[b]
  \centering
  \resizebox{1.0\textwidth}{!}
   {
    \includegraphics{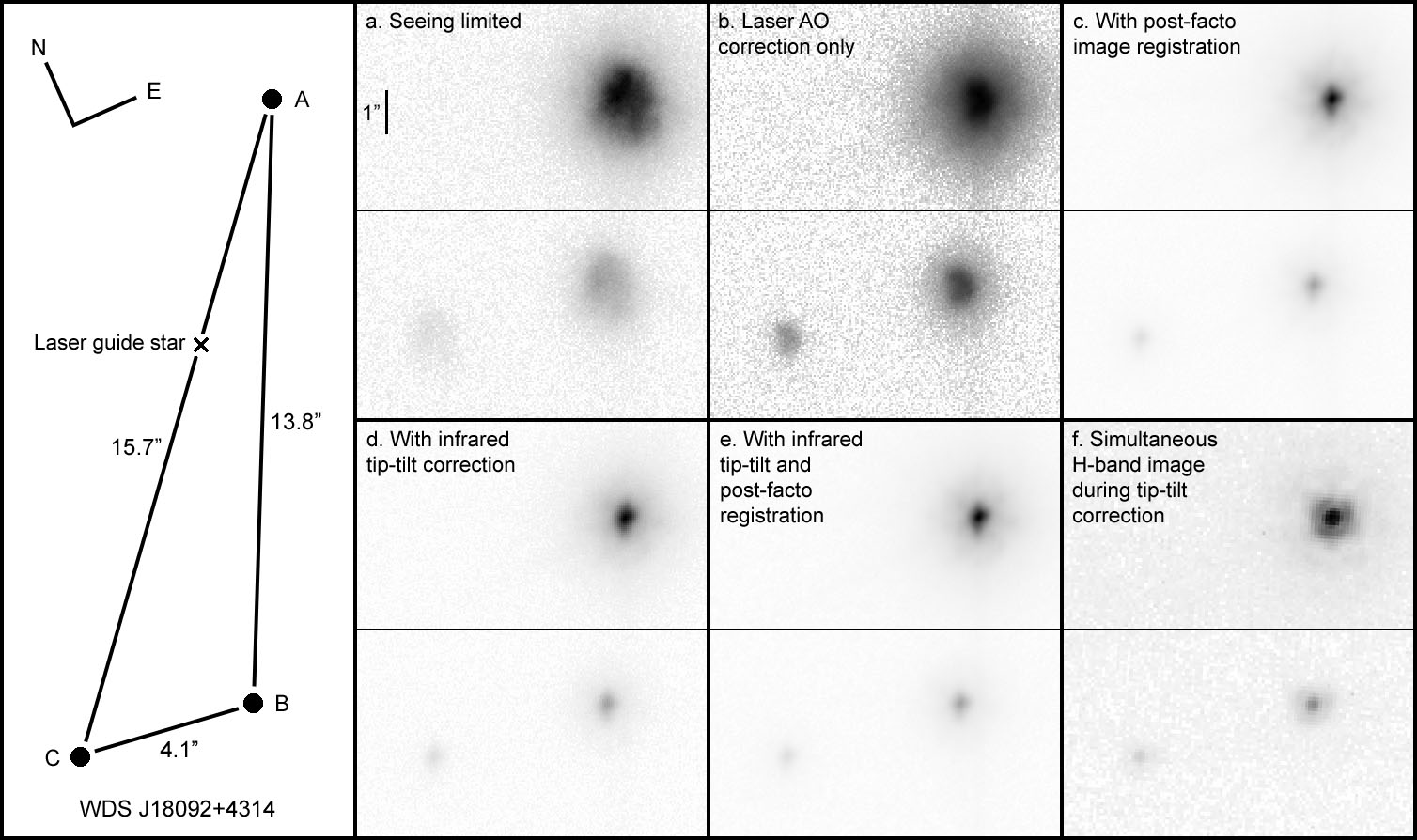}
   }
   \caption{Images of the WDS J18092+4314 triple star system taken with Robo-AO in i' band unless otherwise noted. The left panel shows the geometry and orientation of the stars. The right six panels show cropped images with different adaptive optics and post-facto registration modes being used. Linear scaling is used for i'-band images, square-root scaling is used for H-band images.\label{fig:one}}

\end{figure*}

We observed the triple star system WDS J18092+4314, where the A component has a brightness of $m_V$ = 9.2 \citep{Hog2000} and $m_H$ =7.9 \citep{Cutri2003}. Figure \ref{fig:one} shows the star system geometry along with a series of observations performed with different parts of the adaptive optics system in operation. During the first observation with the main EMCCD camera, the deformable mirror was set to correct for static error only in the telescope in order to measure natural seeing (a). The second observation entailed observing with the EMCCD camera while the high-order adaptive optics correction was enabled. We present images from the EMCCD camera that are simply co-added (b; laser AO correction only) as well as images that have been processed with our standard post-facto image registration techniques (c; e.g., \citealt{Law2014b}). The third observation was taken with simultaneous operation of the high-order adaptive optics loop and the infrared tip-tilt correction. Again we present images from the EMCCD camera that are co-added (d), as well as with post-facto image registration (e).  During this last observation with the infrared tip-tilt correction, full frames from the SAPHIRA were also recorded and co-added (f).

Table \ref{table} presents the measured image metrics from each of the observations. The image with high-order correction and post-facto registration is representative of the typical performance achieved with the Robo-AO system \citep{Baranec2014}. When using infrared tip-tilt correction, the achieved Strehl ratio in the visible was identical to the standard performance within the measurement precision; additional post-facto registration of these images marginally improved the achieved image quality. The achieved Strehl ratio in the infrared of the tip-tilt star, A, 32$\%$, was lower than the second brightest star in the field, B, 56$\%$. Upon inspection of the raw SAPHIRA frames, it was discovered that the tip-tilt star was saturating for approximately the last 10$\%$ of frames before a reset, affecting the ultimate image and adding noise to the tip-tilt correction. Using the unsaturated frames, we calculate a Strehl ratio for A of 48$\%$, and if we use just the 14 frames after each reset, keeping the peak signal at half of the full-well capacity, we calculate a Strehl ratio for A of 57$\%$.

\begin{table*}[b]
\centering
\caption{Image metrics from observations of WDS J18092+4314 in figure \ref{fig:one}.}
\label{table}
\begin{tabular}{lcccccc}
\hline
\multicolumn{1}{c}{{\bf WDS J18092+4314}}          & \multicolumn{2}{c}{{\bf A}}   & \multicolumn{2}{c}{{\bf B}} & \multicolumn{2}{c}{{\bf C}} \\ \hline
Observing mode                                     & SR                & FWHM      & SR          & FWHM          & SR          & FWHM          \\
                                                   & ($\%$)            & (\arcsec) & ($\%$)      & (\arcsec)     & ($\%$)      & (\arcsec)     \\ \hline
a. Seeing limited                                  & -                 & 1.02      & -           & 0.98          & -           & 1.02          \\
b. Laser AO correction only                        & 4.3               & 0.54      & 4.4         & 0.56          & -           & 0.52          \\
c. With post-facto image registration*             & 10.2              & 0.20      & 8.4         & 0.32          & 8.2         & 0.28          \\
d. With infrared tip-tilt correction               & 9.6               & 0.26      & 8.3         & 0.30          & 9.9         & 0.28          \\
e. With IR tip-tilt and post-facto registration*   & 10.3              & 0.20      & 8.6         & 0.28          & 8.8         & 0.30          \\
f. Infrared observation, SR at $\lambda$ = 1635 nm & 57$^\dagger$ (32) & 0.26      & 56          & 0.26          & -           & -             \\ \hline
\end{tabular}
\newline
\newline
Strehl ratio at $\lambda$ = 765 nm unless noted. - denotes low confidence measurement. * Images are up scaled by a pixel factor of 2 as part of the image registration processing. $^\dagger$ The reported Strehl ratio includes only data frames with less than half full-well capacity. The parenthetical Strehl ratio includes all saturated and non-saturated frames.

\end{table*}

Using the Mar\'echal approximation and propagation of typical systematic errors, we were able to check for consistency of the achieved image correction for star B where there was no detector saturation; the H-band Strehl ratio corresponds to a wavefront error of 199$\pm$16 nm RMS, consistent with the i'-band wavefront error of 192$\pm$4 nm RMS.

\subsection{Analysis of tip-tilt correction}

We analyzed the position of the image of star A on the individual EMCCD camera frames with the high-order AO loop closed and with and without infrared tip-tilt correction. Without active tip-tilt compensation the RMS displacement in orthogonal detector coordinates x and y were 0\farcs284 and 0\farcs184, respectively. Infrared tip-tilt correction reduced the RMS displacement to less than the size of a pixel, 0\farcs033 in x and 0\farcs035 in y. The total tip-tilt tracking error is the root-sum-square (RSS) of the measurement and temporal errors. The signal-to-noise ratio (SNR) of the tip-tilt measurement was 36 which, when propagated through an estimate of the residual tip-tilt error (e.g. \citealt{Hardy1998}; eq. 5.15), should have resulted in a one-dimensional measurement displacement error of 0\farcs004 RMS, negligible compared to the 0\farcs11 diffraction-limited core size in i'-band.

\begin{figure*}
  \centering
  \resizebox{1.0\textwidth}{!}
   {
    \includegraphics{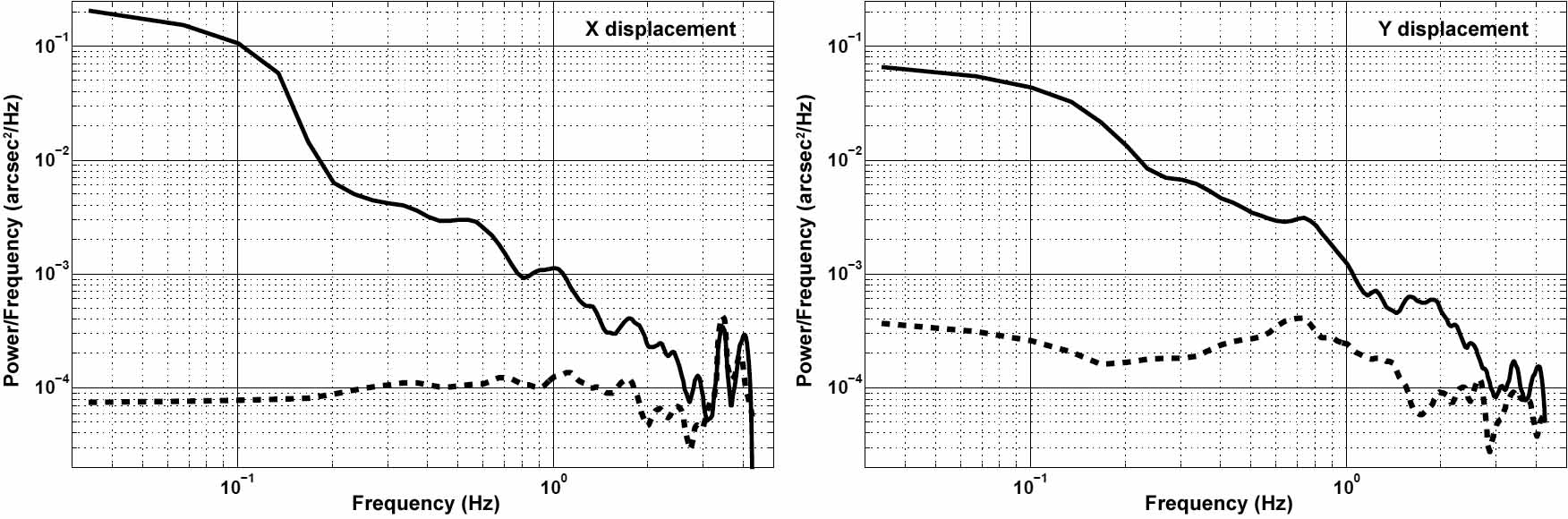}
   }
   \caption{Power spectra of x and y stellar displacement on the visible EMCCD camera with (dashed) and without (solid) infrared tip-tilt correction.\label{fig:two}}

\end{figure*}

The power spectrum of the image position on the EMCCD camera with and without tip-tilt correction is plotted in Figure \ref{fig:two} and shows a crossover rejection frequency of approximately 3 Hz, lower than the 10 Hz expected with a loop rate of 100Hz and gain of 0.5. To investigate this discrepancy we analyzed the calculated center-of-mass position values from the adaptive optics system telemetry. Upon visual comparison of the calculated position with the image in the tip-tilt window we found that, while the position angle was consistent, the algorithm underestimated the magnitude of the stellar displacement from the center of the tip-tilt window. We then calculated the stellar position based on peak tracking via cross-correlation with a Gaussian kernel \citep{Baranec2007} which more closely matched a visual approximation of the stellar position; we found the magnitude of displacement using this method was a factor of 2.35 larger. In practice, this under-calculation of the magnitude of the stellar displacement by using the center-of-mass algorithm lowered our effective loop gain to 0.2 and manifested as a temporal error. When calculating the position of the star on the infrared camera using the more robust centroiding method, we found RMS displacements of 0\farcs068 in both axes. This converts to an effective two-axis image width of 0\farcs25 after convolving with the diffraction-limited core size, closely matching the 0\farcs26 FWHM measured on the EMCCD camera (Table 1, d). When the positions of the star on the infrared camera are binned in time to match the frame rate of the visible camera, essentially a low-pass filter, we found the RMS displacements to be 0\farcs036 and 0\farcs038 in x and y, closely matching the stellar position error observed on the EMCCD camera.

\subsection{Pilot survey of the multiplicity rate of faint M-dwarf stars within 33 pc}
\label{survey}

M-dwarfs are the most common type of star in our galaxy and also the most varied class: they span a factor of six in stellar mass and stellar radius \citep{Leggett1996}. M-dwarf multiplicity properties give windows into stellar formation processes at a wide range of masses and even potentially different formation environments from solar-type stars (e.g., \citealt{Reipurth2014}). As more planets are found around M-dwarfs (e.g., \citealt{Charbonneau2009, Muirhead2012, Bonfils2013}) our understanding of their planetary formation environments will also be informed by their stellar multiplicity properties.

For these reasons, M-dwarfs have been extensively studied by recent high-angular-resolution surveys covering hundreds of targets (e.g., \citealt{Law2008, Bergfors2010, Janson2012, Ziegler2015}), taking advantage of the relatively high time-efficiency of Robo-AO and Lucky Imaging methods. However, these large-sample surveys have been necessarily limited to the higher-mass M-dwarfs because their wavefront sensing (or guide-star-measurement) is conducted in the optical; current large-telescope laser-guide star systems can reach fainter targets but cannot efficiently observe the hundreds-of-targets sample sizes required to perform statistically-significant comparisons across the M-dwarf mass range. Mid-and-late M-dwarfs have extreme optical/NIR colors, reaching $m_R-m_H$=7.5 at M9, compared to $m_R-m_H$ =4.1 at M3 \citep{Kraus2007}. For M-dwarfs later than M5, there are only a few hundred targets across the sky nearby enough for optically-based high-angular-resolution surveys to reach (e.g., \citealt{Law2006, Janson2014}).

Robo-AO is currently engaged in a high-angular-resolution survey of the optically-brightest 3,000 M-dwarfs \citep{Law2014a}. Although we homogeneously cover a much larger sample than previous surveys, allowing a careful comparison of stellar multiplicity properties at varying stellar masses, we need to push to the lowest-mass M-dwarfs to properly cover the entire M-dwarf parameter space. The new infrared capabilities described here give us the ability to address  a much larger sample of late M-dwarfs than would otherwise be possible – covering several thousand more late M-dwarfs \citep{Lepine2005} than can be covered with an optical wavefront sensor.

\begin{figure*}
  \centering
  \resizebox{1.0\textwidth}{!}
   {
    \includegraphics{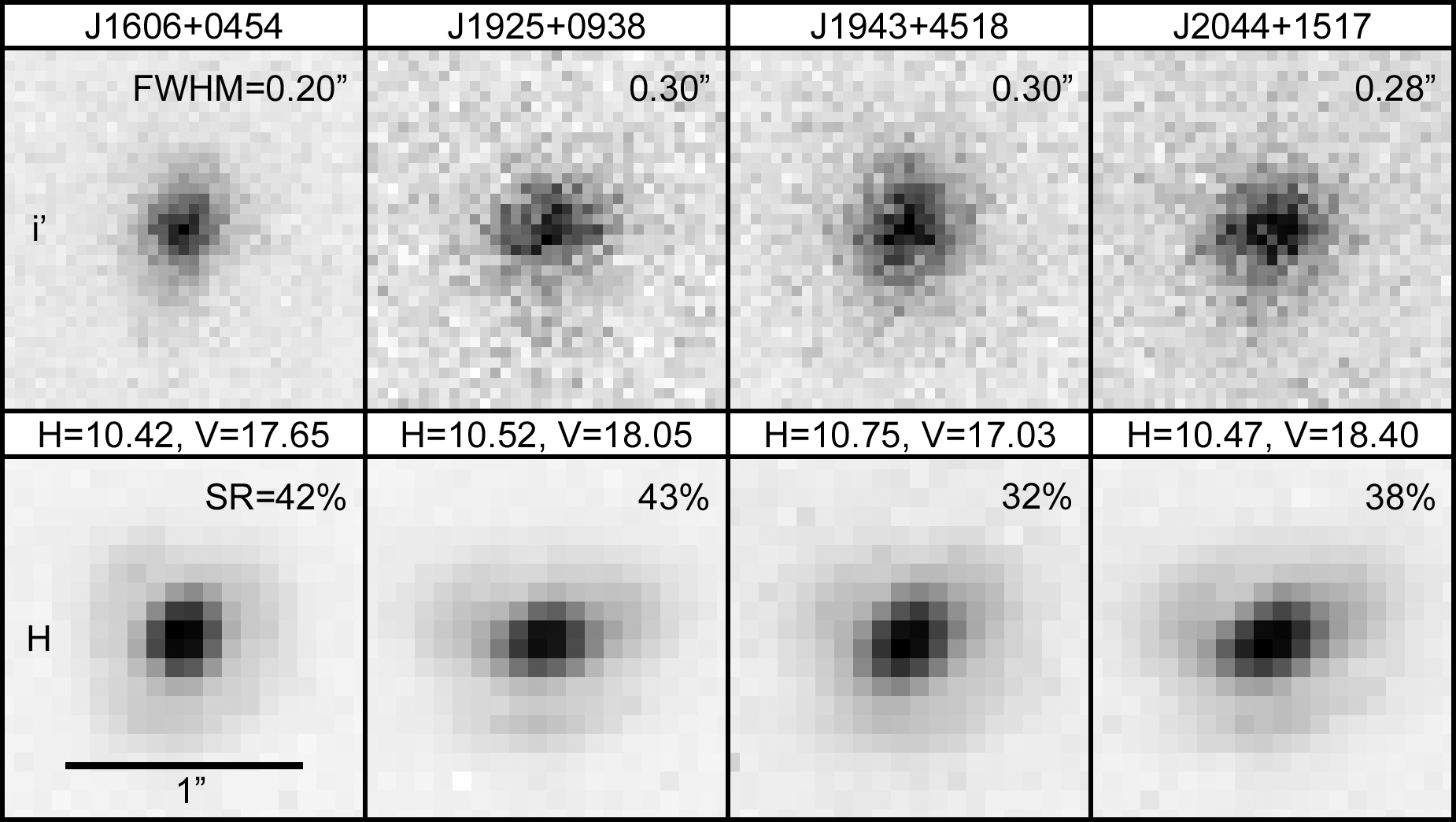}
   }
   \caption{Robo-AO adaptive optics images in the visible (i' band) and infrared (H band) of four M-dwarfs with infrared tip-tilt correction with corresponding image metrics. Each image is 1\farcs5 $\times$ 1\farcs5 and displayed with a linear scale. H magnitudes are from 2MASS All-Sky Point Source Catalog and V magnitudes are from \citealt{Lepine2005}.\label{fig:three}}
   \label{fig:binaries_summary}
\end{figure*}

	To validate our ability to undertake this large survey, we attempted to observe four M-dwarf stars from the Lepine and Shara Proper Motion catalog \citep{Lepine2005} that were otherwise too faint for effective visible-light post facto registration techniques and required the use of infrared tip-tilt sensing -- typical R-band magnitudes of 16-17 and H-band magnitudes of 10-11. Figure \ref{fig:three} shows the resulting images and image metrics from the EMCCD and SAPHIRA camera with all exposures co-added. The achieved Strehl ratio in H band was more modest than for the brighter star in Section \ref{threeone}. Given the very stable seeing, we assume the temporal error to remain the same with any additional error resulting from increased measurement error. The fainter tip-tilt guide sources resulted in a per-frame SNR of $\sim$10 which should only increase the one-dimensional measurement displacement error to 0\farcs013 RMS. We again investigated the calculated stellar center-of-mass position in each frame from telemetry with the cross-correlation method. We found no clear correlation in position angle and the mean difference between the two position calculations over all frames ranged from 0\farcs059 for J1606+0454 to 0\farcs067 for J1943+4518. This additional measurement error accounts for the greater image width and less PSF structure in the visible images presented in Figure~\ref{fig:three}.  Despite this, we were able to achieve visible-light image widths 3-4 times more acute than possible without adaptive optics compensation.  
	
We identified a $\Delta m_{i'}$=1.4 optical companion to J1925+0938 in the EMCCD image that is 5\farcs1 away at a position angle of 134$^{\circ}$. When compared to 2MASS J-band data from 1999, J1925+0938 appears to have changed position compared to all other stars in the field including the optical companion which then appeared to be separated by ~9\arcsec. J1925+0938 has a proper motion of -0\farcs257/year in DEC and +0\farcs075/year in RA (Lepine $\&$ Shara 2005); its new position as well as new angular separation with the optical companion, likely a background object, are consistent with the published proper motion.
\section{Summary and Future Plans}
\label{sec:plans}

We demonstrated the use of a sub-electron read noise infrared avalanche photodiode array as a simultaneous high-speed imaging and tip-tilt wavefront sensing detector and presented preliminary results. We plan to explore the optimization of the infrared tip-tilt control system to achieve improved imaging performance, e.g., by using fewer pixels and guiding just on the core of the stellar PSF, using more robust centroiding routines, and employing more optimal control algorithms (e.g., \citealt{Sivo2014}). In the immediate future, the Robo-AO system will be transferred to the Kitt Peak 2.1-m telescope for a 3 year deployment; we plan to fully integrate an anti-reflection-coated science-grade version of the SAPHIRA detector. An updated robotic queue system will be developed to include both the EMCCD and SAPHIRA cameras with the option to use infrared tip-tilt correction during observations. Subsequently, we intend to execute the multiplicity study of nearby M-dwarfs as presented in Section~\ref{survey}.

Currently we are using the same camera demonstrated here as an upgrade to the infrared speckle nulling camera \citep{Martinache2014} behind the SCExAO system \citep{Jovanovic2014} at Subaru telescope to improve the achievable contrast at infrared wavelengths and to test dark speckle techniques (e.g., \citealt{Labeyrie1995}). SAPHIRA based cameras can also be used to drastically improve the sensitivity of other post-coronagraphic wavefront sensors, e.g., replacing the InGaAs Shack-Hartmann wavefront sensor in the P1640 spectrograph \citep{Cady2013, Vasisht2014} behind the PALM-3000 exoplanet adaptive optics system \citep{Dekany2013} at Palomar Observatory. Additionally we are investigating using the SAPHIRA as an alternative low-order wavefront sensor technology to increase sky coverage at the Keck II telescope, similar to the HAWAII-2RG detector at the Keck I telescope \citep{Wizinowich2014}. We also intend to explore using the SAPHIRA devices as detectors for infrared pyramid wavefront sensors where the $<$ 1e$^-$ read noise will mitigate the need for pixel binning to optimize the spatial sampling of the wavefront for faint targets and where the fast read rates would support extreme adaptive optics.

\acknowledgments

We thank the staff of Palomar Observatory for their support of the infrared camera with the Robo-AO system on the 1.5-m telescope. Development and characterization of the SAPHIRA detectors at the University of Hawai`i is sponsored by the National Science Foundation under Grant No. AST-1106391 and by NASA ROSES APRA award $\#$NNX 13AC13G. The Robo-AO system was developed by collaborating partner institutions, the California Institute of Technology and the Inter-University Centre for Astronomy and Astrophysics, and supported by the National Science Foundation under Grant Nos. AST-0906060, AST-0960343 and AST-1207891, the Mt. Cuba Astronomical Foundation and by a gift from Samuel Oschin. Ongoing science operation support of Robo-AO is provided by the California Institute of Technology and the University of Hawai`i. C.B. acknowledges support from the Alfred P. Sloan Foundation. D.A. is supported by a NASA Space Technology Research Fellowship, grant $\#$NNX 13AL75H. This research has made use of the VizieR catalogue access tool, CDS, Strasbourg, France.

{\it Facility:} \facility{PO:1.5m (Robo-AO)}

\bibliographystyle{apj.bst}
\bibliography{references.bib}

\end{document}